\documentclass[a4paper,12pt]{article}

\usepackage{amsmath}%
\usepackage{amsfonts}%
\usepackage{amssymb}%
\usepackage{graphics} 
\usepackage{graphicx} 

\numberwithin{equation}{section}

\DeclareMathOperator{\Ai}{Ai}
\DeclareMathOperator{\Sh}{Sh}
\DeclareMathOperator{\e}{e}

\newcommand{\abs}[1]{\left\vert#1\right\vert}

\newcommand{\eps}{\varepsilon}

\newcommand{\A}{\mathcal{A}}
\begin{document}
 
\title{Stationary subsonic boundary layer in the regions of local heating of surface}
 
\date {} 
\author{M.V. Koroteev$^{*}$, I.I. Lipatov\thanks{Central Aerohydrodynamic Institute(TsAGI), Zhukovsky, Russia. E-mail: m.koroteev@gmail.com}}
\maketitle

\begin{abstract}
The problem of disturbed flow in the laminar boundary layer is investigated when a heating element is located on the surface of the body. The flow is supposed to be laminar one. It is shown that the problem may be solved in terms of free interaction theory. The solution of linear flat problem is constructed. The results of asymptotic analysis are also presented.
\end{abstract}

\section{Introduction} 
\noindent

The problems of boundary layer control remain important being connected with e.g., the increase of fuel prices. The reduction of vehicle resistance accomplished in particular by means of this control arrives at the reduction of fuel consumption and thus, to the increase of the flight effectiveness. Among a number of methods for boundary layer control the method in which heating elements are located on the surface and vary the temperature in the boundary layer, plays the crucial role. The use of these elements is similar, in some sense, to the application of small humps on the surface which also influences the flow in the boundary layer. In this case the problem of influence of local heating of the surface on the flow in the boundary layer appears. The crucial characteristics in this problem are pressure and shear-stress behaviour. The perturbation of these function in the neighbourhood of the heating element determines how to influence the boundary layer to delay laminar-turbulent transition, reduce the resistance etc. The general consideration of the problem implies the construction of solutions of the nonlinear boundary problems and investigation of perturbations depending on time. However, the approach on the current stage requires to consider simpler stationary models and can be linearized. This approach enables to state the general mechanisms of the flow adjacent to the heated region and estimate the physical effects of action of such heating on the flow.

\vspace{0.5cm}
\section{Scales evaluation and problem statement for the viscous sublayer}
\noindent

We consider the subsonic or supersonic\footnote{We consider the subsonic problem but the discussion in the current section is general both for subsonic and for supersonic case.} compressible flow over the semi infinite flat plate when the flow still may be considered laminar but Reinolds number is sufficiently large. Denote $Re=\epsilon^{-2}$. Thus, $\epsilon$ is assumed to be small parameter. In general, the flow is supposed to be two dimensional and nonstationary one. Preliminary, let us transform the original Navier-Stokes equations to nondimensional ones. Introduce the appropriate nondimensinal variables denoted with the same letters $xl,yl,zl,lu_{\infty}^{-1}t$, $u_{\infty}u,u_{\infty}v,\rho_{\infty}\rho$, $u_{\infty}^{2}R^{-1}T$, $\mu_{\infty}\mu$, $\rho_{\infty}u_{\infty}^{2}p$. Here $u_{\infty},\rho_{\infty},\mu_{\infty}$ are a longitudinal velocity, density and the coefficient of viscosity in the undisturbed flow over the surface where Euler equations are valid, $l$, distance from the leading edge to the place on the plate where the heating region is located. Substituting these functions in Navier-Stokes equations, we obtain the nondimensional equations differing from the former ones only by the multiplier $1/Re$ in the right sides of the equations.

In the problem under consideration there are three different regions with the specific equations. These regions naturally appear while asymptotic expanding of solutions of the Navier-Stokes equations and balancing the orders of the convective, dissipative and pressure terms. This structure is usually called triple-deck. The viscous sublayer on the bottom of the boundary layer is called inner deck, the boundary layer itself and the inviscid flow are called the main deck and the outer deck respectively. We will use these terms interchangeably with the terms the viscous sublayer, boundary layer and inviscid flow.

In previous works the problem of temperature and catalicity jumps has been systematically investigated\cite{1,2}. In this case we assume that the temperature disturbances appear in a region with finite length to be estimated. It was shown in\cite{1,2} that in the problem with the temperature jump $\Delta T\sim T\sim O(1)$ and the disturbances propagate on the distance $\Delta x\sim\epsilon^{3/4}$. Therefore we assume that the order of the length of the heating region is the same as that of the region of disturbances propagation i.e., $a\sim\epsilon^{3/4}$ where $a$ is the length of heated region. It was shown in \cite{3} that this situation leads to forming of a local effective irregularity(hump) on the surface and the problem becomes similar to that of compressible flow over the local irregularities\cite{4,5,6, 7}. Consider the region where the convective, dissipative, and pressure terms of Navier-Stokes equations are of the same order. The estimations of the transversal size of the interaction region follow from the asymptotic equivalence of the convective and dissipative terms
$$
u\frac{\partial u} {\partial x}\sim \epsilon^{2}\frac{\partial^{2}u} {\partial y^{2}}. 
$$
Taking into account that in the sublayer(inner deck) $u\sim y/\epsilon$ we obtain the estimate $y\sim\epsilon^{5/4}$.
Similarly, the asymptotic relation

$$
\frac{\partial p}{\partial x}\sim \epsilon^{2}\frac{\partial^{2} u}{\partial y^{2}} 
$$
leads to the estimate of the pressure in the inner deck $p\sim\epsilon^{1/2}$. Time changing scale is represented by $\tau\sim\eps^{-1/2}$.

The above estimates show that under these circumstances free interaction mode is valid\cite{8, 9, 10}. Asymptotic representations in the regions located above the inner deck were derived in\cite{3}. Consider in detail the boundary problem in the viscous asymptotic sublayer of the nonlinear disturbances which is located beneath the asymptotic Prandtl boundary layer and outer inviscid flow. The functions in the sublayer denoted by the index $3$ have the following asymptotic expansions
$$
x=1+ax_{3}, \quad y=a^{1/3}\epsilon y_{3}, \quad t=\tau t_{3}
$$
$$
u=a^{1/3}u_{3}+\ldots,\quad v=\epsilon a^{-1/3}v_{3}+\ldots, \quad 
$$
$$
p=p_{\infty}+a^{2/3}p_{3}+\ldots, \quad T=T_{3}+\ldots, \quad \rho=\rho_{3}+\ldots, 
$$
where $a$ is the longitudinal scale estimate for the heating region. Substituting the above expansions in Navier-Stokes equations and letting $\epsilon\to 0$ we arrive at the set of equations for the viscous nonstationaty layer which is two dimensional nonstationary boundary layer
$$
\Sh\frac{\partial\rho_{3}}{\partial t_{3}}+\frac{\partial \rho_{3} u_{3}}{\partial x_{3}}+\frac{\partial \rho_{3} v_{3}}{\partial y_{3}}=0 
$$
$$
\rho_{3}\left(\Sh\frac{\partial u_{3}}{\partial t_{3}} +  u_{3}\frac{\partial u_{3}}{\partial x_{3}}+v_{3}\frac{\partial u_{3}}{\partial y_{3}}\right)+ \frac{\partial p_{3}}{\partial x_{3}} = \frac{\partial }{\partial y_{3}}\left(\mu\frac{\partial u_{3}}{\partial y_{3}}\right)
$$
\begin{equation} 
\label{21} 
\end{equation} 
$$
\rho_{3}\left(\Sh\frac{\partial T_{3}}{\partial t_{3}} +  u_{3}\frac{\partial T_{3}}{\partial x_{3}}+v_{3}\frac{\partial T_{3}}{\partial y_{3}}\right) = \frac{\partial }{\partial y_{3}}\left(\frac{\mu}{\Pr}\frac{\partial T_{3}}{\partial y_{3}}\right)
$$ 
$$
\frac{\partial p_{3}}{\partial y_{3}}=0, \quad\rho_{3}T_{3}=1, 
$$
In these equations $\Pr$ is Prandtl number, $\Sh=\tau^{-1}\eps^{1/2}$, Struhal number. The boundary conditions are formulated on the basis of no slip conditions and matching the solutions with these in other regions\cite{2}. For the pressure the interaction condition holds having the following form for the subsonic case\cite{3} 
$$
p_{3} = -\frac{B_{1}}{\pi\sqrt{1-M^{2}}}\int\limits_{-\infty}^{+\infty}\frac{\partial D/\partial\xi}{\xi - x_{3}}d\xi, 
$$
where $D$ is an arbitrary function. The initial conditions are derived by matching with the wall adjacent part of the undisturbed boundary layer on the plate 
$$
u_{3}\to Ay_{3}, T_{3}\to T_{20}(y_{3}), p_{3},D\to 0  
$$
as $x_{3}\to -\infty$. On the surface of the plate $y_{3}=0$ and the no slip conditions yield $u_{3}=v_{3}=0, T_{3}=T_{3w}$. At a large  distance from the plate $y_{3}\to\infty$ and the boundary conditions have the form $u_{3}\to A(y_{3}+D),T_{3}\to T_{20}(0)$, where $T_{20}$ is the main term of temperature asymptotic expansion in the main deck. Let us introduce the additional transform of the independent and dependent variables to simplify the numerical analysis of the problem
$$
x_{3} = x_{b}, \quad y_{3} = A^{-1/3}\int_{0}^{y_{b}}\frac{dy_{b}}{\rho_{b}}, 
$$ 

$$
u_{3}=A^{2/3}u_{b}, 
$$

$$
p_{3}=\rho_{20}(0)A^{2/3}p_{b}, \quad \rho_{3}=\rho_{20}(0)\rho_{b},\quad T_{3} = T_{20}(0)T_{b} 
$$

$$
B_{2}=B_{1}\pi^{-1}\rho_{20}^{-1}(0)A^{-5/3} (1-M^{2})^{-1/2}, \quad  D=A^{-1/3}d 
$$
The value $v_{b}$ is established depending on the new variables. In addition, set $\Pr=const$ and $\rho_{3}\mu=1$ implying that the coefficient of viscosity is linearly dependent on the temperature. Thus, the boundary problem (\ref{21}) is rewritten as follows
$$
\frac{\partial u_{b}}{\partial x_{b}}+\frac{\partial v_{b}}{\partial y_{b}}=0,
$$
$$
\Sh\frac{\partial u_{b}}{\partial t_{b}} + u_{b}\frac{\partial u_{b}}{\partial x_{b}}+v_{b}\frac{\partial u_{b}}{\partial y_{b}}+T_{b}\frac{\partial p_{b}}{\partial x_{b}}=\frac{\partial^{2} u_{b}}{\partial y_{b}^{2}}
$$
\begin{equation}
 \label{2}
\end{equation}
$$
\Sh\frac{\partial T_{b}}{\partial t_{b}} + u_{b}\frac{\partial T_{b}}{\partial x_{b}}+v_{b}\frac{\partial T_{b}}{\partial y_{b}}=\frac{\partial^{2} T_{b}}{\partial y_{b}^{2}}
$$
with the boundary conditions
$$
u_{b} (x_{b},0,t_{b})  =  v_{b} (x_{b},0,t_{b}) = 0,
$$
$$
T_{b} (x_{b},0,t_{b})  =  T_{w}(x_{b},t_{b}) ,
$$
$$
u_{b}\to y_{b}+d, \quad d_{1}=\int_{0}^{+\infty}(1-T_{b})d\eta + d, \quad T_{b}(x,\infty)\to 1,y_{b}\to\infty
$$
$$
d(-\infty)=0, \quad p_{b}(x_{b},t_{b}) = -B_{2}\int\limits_{-\infty}^{+\infty}\frac{\partial d_{1}/\partial\zeta}{\zeta - x_{b}}d\zeta
$$
Note that the interaction condition can be transformed into the other form as\cite{5}
$$
\A^{\prime\prime}_{1b}(x_{b}) = -\int\limits_{-\infty}^{+\infty}\frac{p^{\prime}_{b}(\xi)}{x_{b}-\xi}d\xi
$$
where
$$
\A_{1b}(x_{b})=\int_{0}^{+\infty}(1-T_{b})d\eta + \A_{b}, u_{b}\to y_{b}+\A_{b}(x_{b}), y_{b}\to\infty.
$$
The last relation we will use in the following sections.

\vspace{1cm}
\section{Construction of the solution of flat linearized stationary problem}

\subsection{General construction of the solution for the case of arbitrary temperature in the viscous sublayer}
\vspace{0.5cm}
\noindent

Let us consider the stationary problem. Then (\ref{2}) may be transformed as follows
$$
\frac{\partial u_{b}}{\partial x_{b}}+\frac{\partial v_{b}}{\partial y_{b}} = 0,
$$ 
\begin{equation} 
\label{31} 
u_{b}\frac{\partial u_{b}}{\partial x_{b}}+ v_{b}\frac{\partial u_{b}}{\partial y_{b}} + T_{b}\frac{\partial p_{b}}{\partial x_{b}} = \frac{\partial^{2} u_{b}}{\partial y_{b}^{2}} 
\end{equation} 
$$
u_{b}\frac{\partial T_{b}}{\partial x_{b}}+ v_{b}\frac{\partial T_{b}}{\partial y_{b}} = \frac{\partial^{2} T_{b}}{\partial y_{b}^{2}}
$$
Assume that the temperature disturbances are small i.e., $\Delta T << T\sim O(1)$. Then the solution of (\ref{31}) may be constructed by expanding the flow functions in the main undisturbed part of the boundary layer. Let us introduce the small parameter $\lambda=\Delta T$ and represent the flow functions as follows
$$
u_{b} = y_{b} + \lambda U +O(\lambda^{2}), \quad v_{b}=\lambda V+O(\lambda^{2}),\quad \A_{1b}=\lambda\A_{1}+O(\lambda^{2})
$$ 
$$
T_{b} = 1 + \lambda T+O(\lambda^{2}), \quad p_{b}=\lambda P +O(\lambda^{2}),\quad \A_{b} = \lambda\A +O(\lambda^{2})
$$
as $\lambda\to 0$.
Substitution of these expansions in (\ref{31}) and boundary condition and equalization of the terms with the same powers of $\lambda$ yield the set
$$
\frac{\partial U}{\partial x_{b}}+\frac{\partial  V}{\partial y_{b}} = 0,
$$
$$
y_{b}\frac{\partial U}{\partial x_{b}} +  V  + \frac{\partial P}{\partial x_{b}} = \frac{\partial^{2} U}{\partial y_{b}^{2}}
$$
$$
y_{b}\frac{\partial T}{\partial x_{b}}  =  \frac{\partial^{2} T}{\partial y_{b}^{2}} 
$$
The boundary conditions take the form
$$
U(x_{b},0) = V (x_{b},0)=0 ,T(x_{b},0)  =  T_{w}(x_{b}) ,
$$
$$
U\to \A(x_{b}), \quad \A_{1}(x_{b})=\int_{0}^{+\infty}T d\eta + \A(x_{b}), \quad T(x_{b},\infty)\to 0,y_{b}\to\infty
$$
$$
\A(-\infty)=0, \quad \A^{\prime\prime}_{1}(x_{b}) = -\int\limits_{-\infty}^{+\infty}\frac{P^{\prime}(\xi)}{x_{b}-\xi}d\xi
$$

Applying Fourier transform with respect to $x_{b}$ to the last set, we obtain 
$$
i\xi\tilde{U}+\tilde{V}^{\prime} = 0,
$$
\begin{equation} 
\label{32}
i\xi y_{b}\tilde{U} + \tilde{V} + i\xi\tilde{P} = \tilde{U}^{\prime\prime},
\end{equation}
$$
i\xi y_{b}\tilde{T} = \tilde{T}^{\prime\prime}
$$
Consider the last equation of (\ref{32}). Set $Y = (i\xi)^{1/3}y_{b}$ and $y_{b}>0$. Then the equation becomes
$$
\tilde{T}^{\prime\prime}  =  Y\tilde{T},
$$
where primes denote differentiating with respect to $Y$. 
Thus, in the result of variable changing the variable $Y$ is complex and the equation for Fourier image is Airy equation. To determine the values of $Y$ uniquely it is necessary to specify the branch of multi-valued function $(i\xi)^{1/3}$. Let us consider the complex plane $\omega=\eta+i\xi$. We find for Airy function $\Ai(\omega^{1/3}y_{b})$. Cut the plane $\omega$ along the negative part of real half axis. In the plane with the cut the function $\omega^{1/3}$ allows to specify regular branches. Then take the branch of cubic root such that at $\omega=1$ we have $\omega^{1/3}=1$. In the following consideration we will use the consequence of previous procedure, that for the branch under consideration $\Ai((i\xi)^{1/3}y_{b})=\Ai(\xi^{1/3}y_{b}\e^{i\pi/6})$ as $\xi>0$ and $\Ai((i\xi)^{1/3}y_{b})=\Ai(\xi^{1/3}y_{b}\e^{-i\pi/6})$ as $\xi<0$. 

The boundary condition at infinity is derived from those for (\ref{2}) and has the form $\tilde{T}(\infty) = 0$.
The second boundary condition is set using the temperature distribution on the wall and taking into account its Fourier transform and has the form $\tilde{T}(\xi)=\tilde{f}(\xi)$. Using the new variable $Y$ the equation for Fourier images of temperature becomes Airy equation. The boundary condition at infinity shows the solution of Airy equation must be chosen recessive at infinity. This solution depends on $Y$ complex plane segment under consideration. However, the above suggested specification of the branch shows that in the case under consideration the argument of Airy functions is located in the segment defined by $\abs(arg Y)<\pi/3$ where the recessive solution is represented by means of $Ai(Y)$ and hence the solution is given by 
$$
\tilde{T}(\xi,y_{b})=K(\xi)\Ai(Y).
$$
The boundary condition in zero point gives
$$
K(\xi) = \frac{\tilde{f}(\xi)}{\Ai(0)},
$$
and we obtain the solution for the images of the temperature.

Consider two other equations of (\ref{32}). Differentiating the second one with respect to $y_{b}$ and substituting the result in the first equation we obtain
$$
\tilde{U}^{\prime\prime\prime}=-i\xi y_{b}\tilde{U}^{\prime}.
$$
Introducing notation $\tilde{U}^{\prime}=F$ and changing variables as above, i.e. $Y = (i\xi)^{1/3}y_{b}$, we arrive at Airy equation
$$
F^{\prime\prime}=YF.
$$
In the last equation the primes imply differentiation with respect to $Y$. The boundary condition for this equation follows from the asymptotic relation for the longitudinal velocity as $y_{b}\to+\infty$. The solution of the equation satisfying the boundary condition is represented by Airy function $\Ai(Y)$. The solution also depends on the complex plane segments as in the case of the temperature equation and we do not repeat the above analysis. Notice that the velocity distributions are not crucial functions in the problem under consideration. It is necessary to obtain the wall-shear and the pressure in the viscous sublayer. 
Thus, we find
\begin{equation}
\label{33}
\tilde{U}^{\prime}(\xi,y_{b})=M(\xi)\Ai(Y).
\end{equation}

To determine $M(\xi)$ we observe that the longitudinal impulse equation yields at $y_{b}=0$ 
$$
\frac{\partial{P}}{\partial{x_{b}}}=\frac{\partial^{2}{U}}{\partial{y_{b}^{2}}}(x_{b},0).
$$
For the pressure derivative we denote $P^{\prime}_{x_{b}}=G(x_{b})$. Then the equation for images takes the form
\begin{equation}
\label{34}
\tilde{G}=\tilde{U}^{\prime\prime}(\xi,0)
\end{equation}
Differentiating (\ref{33}) with respect to $y_{b}$, evaluating it at $y_{b}=0$ and substituting in (\ref{34}), we have
$$
\tilde{G}=M(\xi)(i\xi)^{1/3}\Ai^{\prime}(0).
$$
The second necessary relation for evaluating $\tilde{G}$ and $M(\xi)$ follows from the interaction condition and has the form
$$
U^{\prime\prime}(x_{b},\infty)+\int\limits_{0}^{+\infty}T^{\prime\prime}d\eta = -\int\limits_{-\infty}^{+\infty}\frac{P^{\prime}(\zeta)}{x_{b}-\zeta}d\zeta
$$
Applying the Fourier transform to the last equality, we obtain
$$
(i\xi)^{2}\tilde{U}(\xi,\infty)+(i\xi)^{2}\int\limits_{0}^{+\infty}\tilde{T}(\xi,\eta)d\eta =-\int\limits_{-\infty}^{+\infty}\e^{-i\xi x_{b}}dx_{b}\int\limits_{-\infty}^{+\infty}\frac{P^{\prime}(\zeta)}{x_{b}-\zeta}d\zeta=
$$
$$
=-i\pi\frac{\abs{\xi}}{\xi}\int\limits_{-\infty}^{+\infty}\e^{-i\xi\zeta}P^{\prime}(\zeta)d\zeta = i\pi\frac{\abs{\xi}}{\xi}\hat{G}
$$
Next, observe that from the foregoing analysis follows
$$
\tilde{U}(\xi,\infty)=\frac{M(\xi)}{(i\xi)^{1/3}}\int\limits_{0}^{+\infty}\Ai(z)dz, \quad \tilde{T} = \tilde{f}\frac{\Ai(Y)}{\Ai(0)},
$$
we arrive at the following condition
$$
\left(i\pi\frac{\abs{\xi}}{\xi} + \frac{(i\xi)^{4/3}}{k^{4/3}}\right)\tilde{G}=\frac{(i\xi)^{5/3}\tilde{f(\xi)}}{\Ai(0)}\int\limits_{0}^{+\infty}\Ai(z)dz,
$$
where we use the notation $k=(-\Ai^{\prime}(0)/\int_{0}^{+\infty}\Ai(z)dz)^{3/4}=0.827$\cite{12}. Here it is necessary to return to the problem of Airy function argument choosing. The last discussion arrive at improper integral of $\Ai(Y)$ calculation which converges as $\abs{arg(Y)}<\pi/3$\cite{11}. The branch of $\omega^{1/3}$ must be specified to provide the convergence of the integral both as $\xi>0$ and $\xi<0$. The above specified branch provides this convergence.

Taking into account two conditions for $\tilde{G}$ and $M(\xi)$ we obtain
$$
\tilde{G} = -\frac{\Ai^{\prime}(0)}{\Ai(0)}\frac{(i\xi)^{5/3}\tilde{f}(\xi)}{\left(i\pi\frac{\abs{\xi}}{\xi}k^{4/3}+(i\xi)^{4/3}\right)}
$$
$$
M(\xi) = -\frac{1}{\Ai(0)}\frac{(i\xi)^{4/3}\tilde{f}(\xi)}{\left(i\pi\frac{\abs{\xi}}{\xi}k^{4/3}+(i\xi)^{4/3}\right)}.
$$
Lastly, for the Fourier images of the shear we find from the previous relation
$$
\tilde{\tau}=\tilde{U}^{\prime}_{y_{b}}(\xi,0) = M(\xi)\Ai(0) = -\frac{(i\xi)^{4/3}\tilde{f}(\xi)}{\left(i\pi\frac{\abs{\xi}}{\xi}k^{4/3}+(i\xi)^{4/3}\right)}.
$$
In all above formulas the plus is chosen at $\xi>0$ and the minus at $\xi<0$.

\subsection{Construction of the solution and asymptotic \\estimations of flow functions}
\vspace{0.5cm}
\noindent

The relations for the functions $M(\xi)$ and $\tilde{G}$ show that in the case under consideration the influence of the function $\tilde{f}(\xi)$ on convergence of integrals for the wall-shear and the pressure is more significant than in the supersonic case\cite{13}. It follows that the shape of local heating region has more influence on the functions in the lower-deck than in the supersonic case. Let us note the properties of $\tilde{f}(\xi)$ to be used below. Since we consider the local regions of heating, we assume this function to be Fourier transform of some finite function. Then, the original finite function is assumed to be even, implying its Fourier transform to be a real function. We consider the pressure case in more detail for the below analysis is similar for the wall-shear.

The pressure gradient on the wall is given by
$$
P^{\prime}_{x_{b}}=G(x_{b}) = -\frac{1}{2\pi}\int\limits_{-\infty}^{+\infty}\tilde{G}(\xi)\e^{i\xi x_{b}}d\xi=
$$
$$
 = -\frac{1}{2\pi}\frac{\Ai^{\prime}(0)}{\Ai(0)}\left[\int\limits_{-\infty}^{0}\frac{(i\xi)^{5/3}\tilde{f}(\xi)}{-i\pi k^{4/3}+(i\xi)^{4/3}}\e^{i\xi x_{b}}d\xi+
\int\limits_{0}^{+\infty}\frac{(i\xi)^{5/3}\tilde{f}(\xi)}{i\pi k^{4/3}+(i\xi)^{4/3}}\e^{i\xi x_{b}}d\xi\right].
$$
Accomplishing the simple transformations and using the results of \S 3.1, introducing the variable $r>0$ such that as $\xi>0$ we have $i\xi=r\e^{i\pi/2}$ and as $\xi<0$ $i\xi=r\e^{-i\pi}$, we obtain
$$
G(x_{b})= -\frac{1}{2\pi}\frac{\Ai^{\prime}(0)}{\Ai(0)}\left[\int\limits_{0}^{+\infty}\frac{\e^{- i5\pi/6}r^{5/3}\tilde{f}(-r)}{-i\pi k^{4/3}+\e^{-i2\pi/3}r^{4/3}}\e^{-ir x_{b}}dr\right.+
$$
\begin{equation}
\label{36}
+\left.\int\limits_{0}^{+\infty}\frac{\e^{i5\pi/6}r^{5/3}\tilde{f}(r)}{i\pi k^{4/3}+\e^{i2\pi/3}r^{4/3}}\e^{ir x_{b}}dr\right].
\end{equation}
The heating region of the surface has some length, therefore to estimate the shear and the pressure we set the temperature on the wall by means of the following finite even function
$$
T(x_{b},0)=\left\{
\begin{aligned}
0.2,&\abs{x_{b}}\le 0.5\\
0, & \abs{x_{b}}>0.5.\\
\end{aligned}
\right.
$$
Fourier transform of this function yields 
$$
\tilde{f}(\xi)=\frac{0.4}{\xi}\sin\left(\frac{\xi}{2}\right).
$$
It follows that $\tilde{f}(\xi)\to 0$ as $\xi\to\infty$ and even. In addition, it is easily seen that in (\ref{36}) we obtain the sum of two conjugate functions. Substituting the last relation in (\ref{36}) and taking into account the complex conjugation, we obtain the integral representation of the general solution in the sublayer for the pressure
\begin{equation}
\label{47}
P^{\prime}(x_{b})=-\frac{0.4}{\pi}\frac{\Ai^{\prime}(0)}{\Ai(0)}\int\limits_{0}^{+\infty}\frac{r^{2/3}\sin(r x_{b} - \psi)\sin(r/2)dr}{\sqrt{\pi^{2}k^{8/3} + \sqrt{3}\pi k^{4/3}r^{4/3}+r^{8/3}}},
\end{equation}
where $\psi$ is determined by
$$
\psi = \arctan\left(\frac{\pi k^{4/3}+\sqrt{3}r^{4/3}}{r^{4/3}+\sqrt{3}\pi k^{4/3}}\right)
$$
In the similar way for the shear we obtain
$$
\frac{\partial{U}}{\partial{y_{b}}}(x_{b},0)=-\frac{0.4}{\pi}\int\limits_{0}^{+\infty}\frac{r^{1/3}\sin(r x_{b} - \phi)\sin(r/2)dr}{\sqrt{\pi^{2}k^{8/3} + \sqrt{3}\pi k^{4/3}r^{4/3}+r^{8/3}}},
$$
where
$$
\phi = \arctan\left(\frac{\frac{\sqrt{3}}{2}\pi k^{4/3}+r^{4/3}}{\frac{1}{2}\pi k^{4/3}}\right)
$$
\begin{figure}
\includegraphics[width=400pt]{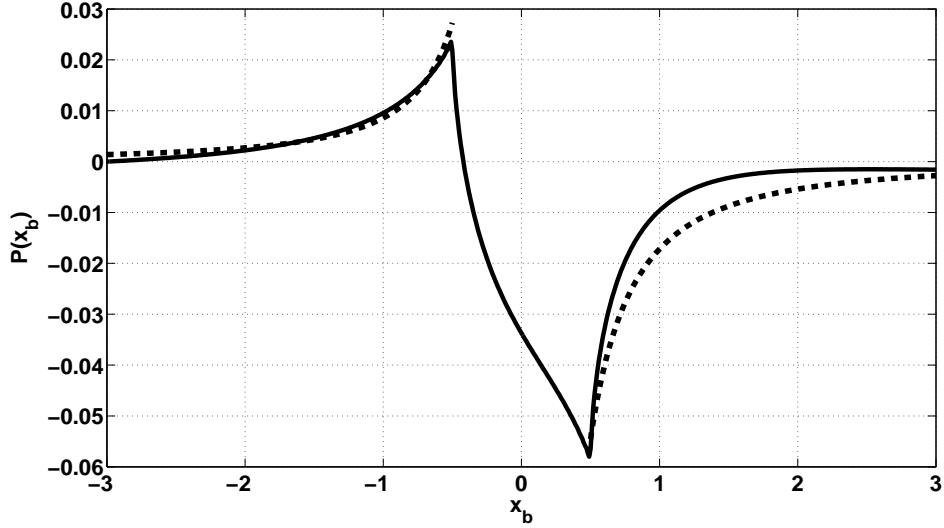}
\caption{Pressure in the viscous sublayer. Dashed lines - asymptotic curves}
\end{figure}
\begin{figure}
\includegraphics[width=400pt]{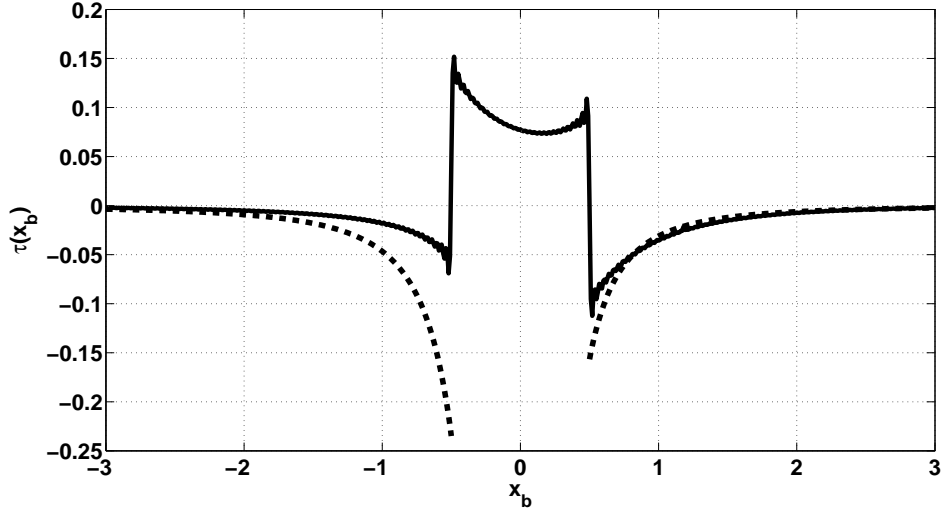}
\caption{Wall-shear in the viscous sublayer. Dashed lines - asymptotic curves}
\end{figure}

The results of pressure estimation are depicted in fig. 1, and the wall-shear, in fig. 2. The pressure estimation is obtained by means of integration of (\ref{47}).

Let us consider the asymptotic behaviour of flow functions and firstly consider the wall-shear case. The main relation we represent as the sum of integrals as follows
$$
\frac{\partial{U}}{\partial{y_{b}}}(x_{b},0)=-\frac{0.4}{\pi}\int\limits_{0}^{+\infty}\frac{r^{1/3}\left(\frac{\sqrt{3}}{2}\pi k^{4/3}+r^{4/3}\right)\sin(r/2)\cos(rx_{b})dr}{\pi^{2}k^{8/3} + \sqrt{3}\pi k^{4/3}r^{4/3}+r^{8/3}}+
$$
$$
+\frac{0.4}{\pi}\int\limits_{0}^{+\infty}\frac{r^{1/3}\frac{1}{2}\pi k^{4/3}\sin(r/2)\sin(rx_{b})dr}{\pi^{2}k^{8/3} + \sqrt{3}\pi k^{4/3}r^{4/3}+r^{8/3}}=
$$
$$
=-\frac{0.4}{\pi}\frac{1}{4i}\left[\int\limits_{0}^{+\infty}r^{1/3}h_{1}(r)\e^{ir(x_{b}+1/2)}dr+\int\limits_{0}^{+\infty}r^{1/3}h_{1}(r)\e^{-ir(x_{b}-1/2)}dr-\right.
$$
$$
-\left.\int\limits_{0}^{+\infty}r^{1/3}h_{1}(r)\e^{ir(x_{b}-1/2)}dr-\int\limits_{0}^{+\infty}r^{1/3}h_{1}(r)\e^{-ir(x_{b}+1/2)}dr\right]-
$$
$$
+\frac{0.4}{\pi}\frac{1}{4}\left[\int\limits_{0}^{+\infty}r^{1/3}h_{2}(r)\e^{ir(x_{b}+1/2)}dr-\int\limits_{0}^{+\infty}r^{1/3}h_{2}(r)\e^{-ir(x_{b}-1/2)}dr-\right.
$$
$$
-\left.\int\limits_{0}^{+\infty}r^{1/3}h_{2}(r)\e^{ir(x_{b}-1/2)}dr+\int\limits_{0}^{+\infty}r^{1/3}h_{2}(r)\e^{-ir(x_{b}+1/2)}dr\right],
$$
where we use the following notation
$$
h_{1}(r)=\frac{\frac{\sqrt{3}}{2}\pi k^{4/3}+r^{4/3}}{\pi^{2}k^{8/3} + \sqrt{3}\pi k^{4/3}r^{4/3}+r^{8/3}},\quad h_{2}(r)=\frac{\frac{1}{2}\pi k^{4/3}}{\pi^{2}k^{8/3} + \sqrt{3}\pi k^{4/3}r^{4/3}+r^{8/3}}.
$$
Observe that
$$
h_{1}(0)=\frac{\sqrt{3}}{2}\pi k^{4/3}, \quad h_{2}(0)=\frac{1}{2}\pi k^{4/3}, \quad h^{\prime}_{1}(0)=h^{\prime}_{2}(0)=0.
$$
For the above integrals the asymptotic relations as $x_{b}\to +\infty$ can be given being the specific case of relations described in \cite{14}, namely
$$
\int\limits_{0}^{+\infty}r^{1/3}h(r)\e^{\pm ir\xi}dr \sim\sum\limits_{k=0}^{\infty}K_{-k-1}(\xi,0)h^{(k)}(0), \xi\to +\infty
$$
where
$$
K_{-k-1}(\xi,0)=\frac{(-1)^{k+1}}{k!}\e^{\pm i\frac{\pi}{2}(k+\frac{4}{3})}\Gamma\left(k+\frac{4}{3}\right)\xi^{-k-\frac{4}{3}},
$$
and represent the integrated kernels of the main kernel $K_{0}(\xi,r)=r^{1/3}\e^{\pi ir\xi}$. Even though the main kernels $K_{0}(\xi,r)$ 
are different for each of the integrals but being of oscillatory type they allow to construct the asymptotic expansion\cite{14}.
Applying the suggested expansion to each of the above integrals, we obtain the following estimation of the downstream wall-shear
$$
\frac{\partial{U}}{\partial{y_{b}}}(x_{b},0)= -\frac{0.8}{9\pi^{2}k^{4/3}}\Gamma\left(\frac{1}{3}\right)\frac{1}{x_{b}^{7/3}} + O(x_{b}^{-7/3})), \quad x_{b}\to+\infty
$$
As $x_{b}\to -\infty$ the slight modification of the analysis yields the following asymptotic relation
$$
\frac{\partial{U}}{\partial{y_{b}}}(x_{b},0)= -\frac{0.4}{3\pi^{2}k^{4/3}}\Gamma\left(\frac{1}{3}\right)\frac{1}{(-x_{b})^{7/3}} + O((-x_{b})^{-7/3})), \quad x_{b}\to -\infty.
$$
Thus, the wall-shear asymptotics upstream and downstream has the same order of vanishing differing this case from the supersonic one. The obtained asymptotic curves are depicted on the fig. 2, and fit well with the precise solution at large $x_{b}$. Note that in the above general asymptotic formula the function $h(r)$ is supposed to be infinite differentiable in zero, which can cause difficulties as e.g., in the case under consideration. However, the main formula relies on the integration by parts and therefore may be applied successively till the smoothness of $h(r)$ allows. Nevertheless, we also note that the smoothness conditions on $h(r)$ in zero may be significantly weakened, not affecting the obtained results.

The asymptotics of the pressure is estimated quite similarly to the case involved and we represent only the final results. Estimating in the similar way for (\ref{47}) we obtain
$$
P^{\prime}(x_{b})= -\frac{2\Gamma\left(\frac{2}{3}\right)}{9\pi^{2}k^{4/3}}\frac{\Ai^{\prime}(0)}{\Ai(0)}\frac{1}{x_{b}^{8/3}}+ o\left(\frac{1}{x_{b}^{8/3}}\right), \quad x_{b}\to +\infty.
$$
Since the integration of this asymptotic expansion is valid\cite{11}, we obtain the estimate
$$
P(x_{b})= \frac{2\Gamma\left(\frac{2}{3}\right)}{15\pi^{2}k^{4/3}}\frac{\Ai^{\prime}(0)}{\Ai(0)}\frac{1}{x_{b}^{5/3}}+ o\left(\frac{1}{x_{b}^{5/3}}\right), \quad x_{b}\to +\infty.
$$
The similar consideration for the case $x_{b}\to -\infty$ gives the estimate
$$
P(x_{b})= -\frac{\Gamma\left(\frac{2}{3}\right)}{15\pi^{2}k^{4/3}}\frac{\Ai^{\prime}(0)}{\Ai(0)}\frac{1}{(-x_{b})^{5/3}}+ o\left(\frac{1}{
(-x_{b})^{5/3}}\right), \quad x_{b}\to -\infty.
$$
On the fig. 1 the curves of asymptotics and the precise solutions of the pressure are depicted.

\section{Discussion}
\noindent

The results show that though qualitatively the flow functions are similar to those for the problems of local irregularities but there are also significant differences. The curves shown in fig. 1,2 allow to conclude that the wall-shear and pressure are described by continuous functions. In addition, the shear in the problem under consideration is similar to that obtained in the problem for the supersonic flow\cite{13}. Downstream the shear increases(fig. 2) implying favourable pressure gradient(fig. 1). The possible points of flow separation are boundaries of heating region upstream and downstream. However, the asymptotic analysis reveals the vanishing of perturbations in the problem is different from that in the supersonic flow. Particularly, for the wall-shear in the supersonic case it was obtained that the perturbations upstream vanish stronger than in the subsonic case.
This fact is easily explained from the physical point of view. The supersonic flow "blows off" the upstream perturbations while in the supersonic case the flow allows the perturbations to propagate upstream. Mathematically this fact can be explained by pole of the integrand for the flow functions in the supersonic case in the lower half plane resulting in exponential vanishing upstream. The similar situation is valid for the pressure. As we mentioned above, the shape of the effective hump in the problem under consideration strongly influences on the pressure in the subsonic case that also can be explained by the fact that the perturbations propagate further upstream. Asymptotic analysis also shows that the obtained representations for the asymptotic behaviour of flow functions fit well with the precise formulas on the distance to be equal to about the half of this region and can be used in the numerical estimations e.g. for the numerical approximation of the interaction condition.

\end{document}